\documentstyle[epsfig]{aipproc}

\newcommand{\gsim}{\raisebox{-4pt}{$\,\stackrel{\textstyle >}{\sim}\,$}}
\newcommand{\lsim}{\raisebox{-4pt}{$\,\stackrel{\textstyle <}{\sim}\,$}}
\newcommand{\phitwopi}{\Phi_{2\pi}}

\begin{document}

\begin{flushright}
WU B 00-18 \\
hep-ph/0010040
\end{flushright}
\vskip 2.0\baselineskip

\title{Two-photon annihilation into pion pairs\thanks{
Talk given at PHOTON2000, Ambleside, England, August 2000;
to appear in the proceedings.} 
}

\author{C. Vogt}
%\thanks{Supported by the Deutsche Forschungsgemeinschaft}}
\address{Fachbereich Physik\\
Universit\"at Wuppertal\\
42097 Wuppertal, Germany}

%\lefthead{LEFT head}
%\righthead{RIGHT head}
\maketitle

\begin{abstract}
We discuss pion pair production in two-photon collisions in two different 
kinematical regimes. When both photons are real and at moderately large 
center-of-mass energy $\sqrt{s}$ we elaborate on partonic transverse momentum
and Sudakov corrections within the hard scattering approach. We also point out
the difference between our approach and that of other authors. When one of the 
photons is highly virtual the produced pion pair can be described in terms of 
a two-pion distribution amplitude, for which we derive the perturbative limit 
at large~$s$.
\end{abstract}

Due to the pointlike structure of the photon exclusive hadron production in
two-photon collisions provides a very useful field for the test of perturbative
QCD. In the limit of large $\sqrt{s}$, the amplitude of 
$\gamma^{(*)}\gamma\to\pi\pi$ factorises into a perturbatively calculable hard 
photon-parton scattering, which in lowest order can simply be obtained from 
one-gluon exchange diagrams, and soft parts that are expressed in terms of 
distribution amplitudes describing the transition of partons to 
pions~\cite{LepageBrodsky1980}. 

At large c.m. energies $\sqrt{s}$, transverse momenta of the partons relative 
to the pion are negligible and the conventional collinear hard scattering 
formula can be applied\cite{BrodskyLepage1981}. At moderately large $\sqrt{s}$
of a few GeV, however, the collinear approach is known to suffer severely
from substantial endpoint contributions where the strong coupling $\alpha_s$
becomes large, such that perturbation theory is not applicable~\cite{isg}. 
These problems can be overcome by including transverse momenta and Sudakov
corrections\cite{BottsSterman,LiSterman}. The correspondingly modified hard 
scattering approach leads to perturbative predictions, which in most cases are
not sufficient to account for the experimental data 
\cite{JakobKroll,Bolz,Stefanis1999}. Hadronic form factors and Compton 
scattering, for example, are dominated by soft contributions at presently 
accessible c.m. energies~\cite{JakobKrollRaulfs,DFJK,CV2000}. 
In this talk we will discuss Sudakov suppressions in 
$\gamma\gamma\to\pi^+\pi^-$ in the few GeV region
and point out the difference between our approach and that of Ref.~\cite{BJPR}.

Another interesting, more theoretically motivated application of the hard
scattering approach is the process $\gamma^*\gamma\to\pi^+\pi^-$ at large
photon virtuality $Q^2$ and large $s$. In this kinematical regime, we will
briefly outline the calculation of the perturbative limit of the two-pion 
distribution amplitude ($2\pi$-DA)~\cite{DFKV}.

\section*{Sudakov suppression in $\gamma\gamma\to\pi^+\pi^-$}
 
\begin{figure}[ht]
\begin{center}
\psfig{file=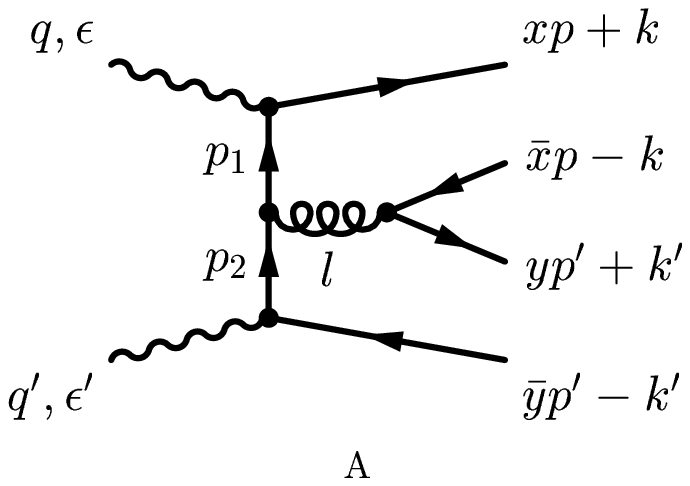,bb=190 640 400 790,width=5cm,angle=0}
\psfig{file=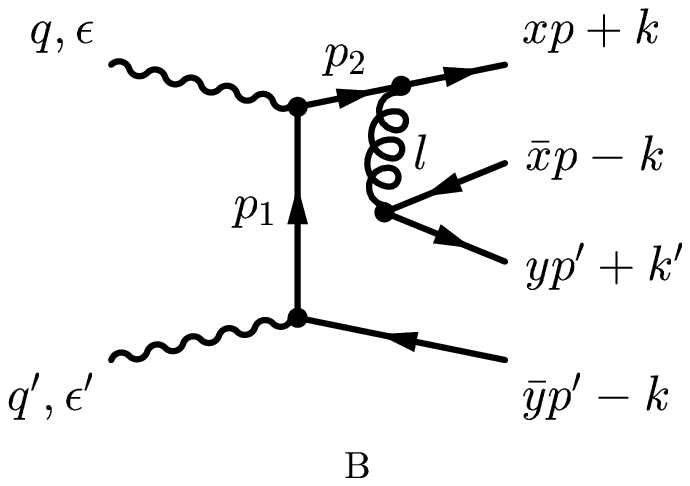,bb=190 640 400 790,width=5cm,angle=0}
\psfig{file=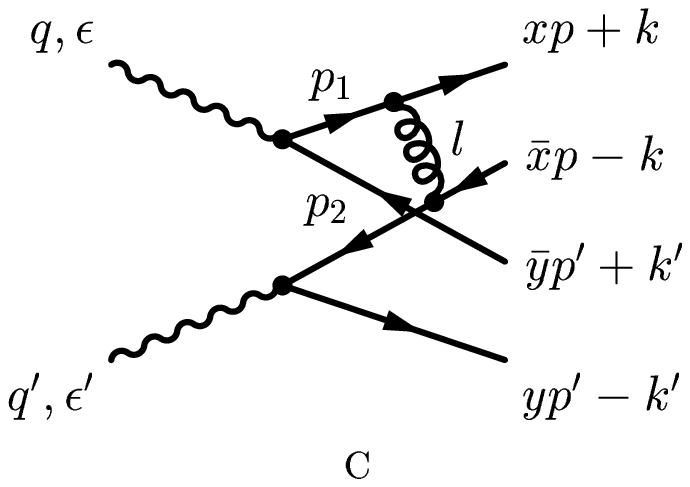,bb=190 640 400 790,width=5cm,angle=0}
\psfig{file=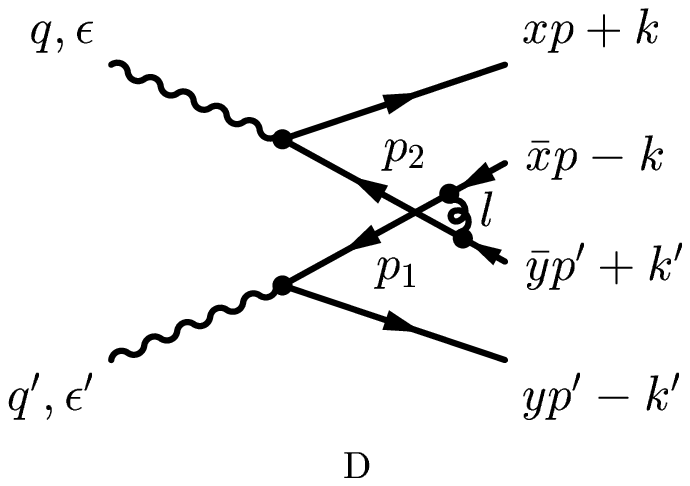,bb=190 640 400 790,width=5cm,angle=0}
\caption{The four basic diagrams for $\gamma\gamma\to\pi\pi$. We use the 
         notation $\bar{x}\equiv 1-x$.}      \label{hsp diagrams}
\end{center}
\end{figure}

In close analogy to the calculation of hadronic form factors 
\cite{LiSterman,JakobKroll,Bolz} in the modified perturbative approach we can 
express the helicity amplitude ${\cal M}_{\lambda\lambda'}$ of the process
$\gamma\gamma\to\pi^+\pi^-$ in transverse configuration space as the 
convolution
\begin{eqnarray}
 {\cal M}_{\lambda\lambda'}(s,\Theta)&=&\int dx\,dy\,
 \frac{d^2 {\bf b}_\perp}{4\pi}\,\frac{d^2 {\bf b}_\perp'}{4\pi}\;\
 \hat{\Psi}_\pi(x,{\bf b}_\perp)\;\hat{\Psi}_\pi(y,{\bf b}'_\perp)\nonumber\\
 &&\times\hat{T}_{H,\lambda\lambda'}(x,y,{\bf b}_\perp,{\bf b}'_\perp;
 s,\Theta,\mu_R)\;\exp[-S(x,y,b_\perp,b'_\perp;\mu_R)], \label{amplitude}
\end{eqnarray}  
where $\Theta$ is the scattering angle in the center-of-mass system of the 
produced pions and $\lambda,\,\lambda'$ are the photon helicities. 
The hat denotes the Fourier transform of a function w.r.t.
the transverse momenta ${\bf k}_\perp,\,{\bf k}'_\perp$ of the partons 
relative to the pions. The Fourier conjugated variables 
${\bf b}_\perp,\,{\bf b}'_\perp$ are the transverse separations of the 
quark-antiquark pairs and $x,\,y$ describe how they share the pions' 
longitudinal momenta. 

Using a phenomenological ansatz for the wave function of the pion's valence 
Fock state we write
\begin{equation}
 \Psi_\pi(x,{\bf k}_\perp)=\frac{\sqrt{6}\,\pi}{f_\pi}
    \exp\bigg[-\frac{k_\perp^2}{8\pi f_\pi^2\,x \, (1-x)}\bigg], \label{lcwf}
\end{equation} 
with $f_\pi=131$ MeV being the pion decay constant. Integrating 
Eq.~(\ref{lcwf}) over transverse momenta leads to the asymptotic form of the 
pion distribution amplitude $\phi_\pi$. The use of the asymptotic form is 
justified through the phenomenology of the $\pi$-$\gamma$-transition form 
factor~\cite{Musatov} and the parameters of expression~(\ref{lcwf}) are fixed 
by various pion decay processes~\cite{BHL}. The Gaussian $k_\perp$-dependence
describes well soft contributions~\cite{JakobKroll,DFJK,CV2000}. 

The Sudakov corrections are incorporated in the factor $e^{-S}$, where $S$ is
the Sudakov function \cite{BottsSterman} (see also \cite{Stefanis2000}). Since
it suppresses large quark-antiquark separations it serves as a natural 
infrared cut-off and thus no external regulator is needed to avoid the
singularity of $\alpha_s$. 

In leading order QCD, the hard photon-parton scattering amplitude $T_H$ is to 
be calculated from 20 one-gluon exchange diagrams, four representatives
of which are shown in Fig.\ref{hsp diagrams}. Following the authors of 
Ref.~\cite{LiSterman} we choose the renormalisation scale $\mu_R$ to be the 
largest mass scale appearing in the gluon virtualities. Owing to the structure
of the hard scattering amplitude its analytical Fourier transform cannot be 
calculated exactly and we have to resort to approximations. As the 
longitudinal momentum fractions occur quadratically in the gluon propagators 
we keep transverse momentum corrections there if not otherwise stated. 

\begin{figure}[t]
\begin{center}
\psfig{file=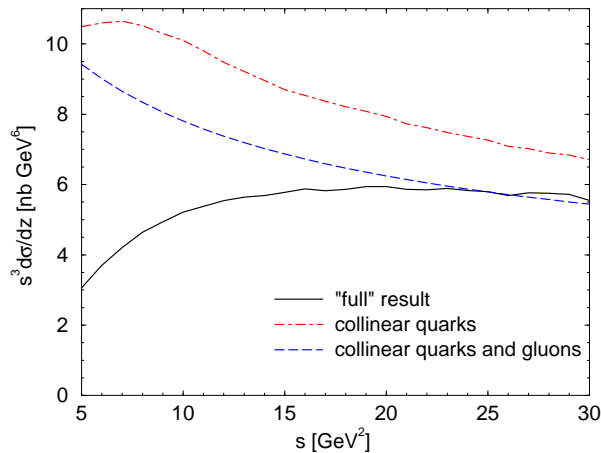,bb=100 70 570 675,width=6cm,angle=-90}
\caption{Differential cross section at c.m. angle $\Theta=90^\circ$ with 
         different approximations.}          \label{dsdz plot}
\end{center}
\end{figure}

Our results for the differential cross section at $\Theta=90^\circ$ using 
different approximations for the quark propagators are shown in 
Fig.\ref{dsdz plot}.
The dot-dashed line shows the result obtained by replacing the quark 
propagators by their collinear limits. In the solid curve we take into 
account transverse momenta in quark propagators in those integration regions
where they have singularities. We see that the effect in the few GeV region
is dramatic, which means that one can generally not ignore 
$k_\perp$-corrections in quark propagators, as has been done in 
Ref.~\cite{Hyer}, for instance. For comparison we also show the result of the
collinear hard scattering approach, i.e. completely neglecting 
$k_\perp$-corrections in quark as well as in gluon propagators, where we have 
frozen $\alpha_s$ below 1 GeV (dashed line). 
We note that our main result, given by the
solid curve, approaches the collinear approximation for $s\gsim20$ GeV$^2$,
i.e. for c.m. energies above 4-5 GeV. In brief, the $k_\perp$-corrections
of the quark propagators effect the transition amplitude such that it
reduces its absolute magnitude while receiving a large phase.

\begin{figure}[t]
\begin{center}
\psfig{file=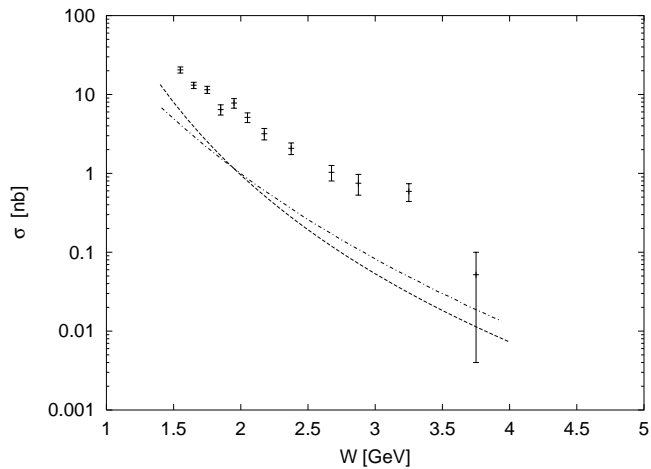,bb=60 60 540 750,width=6cm,angle=-90}
\end{center}
\caption{The combined cross section $\sigma(\gamma\gamma\to\pi^+\pi^-,K^+ K^-)$
         as a function of the c.m. energy $W=\sqrt{s}$.} \label{sigma plot}
\end{figure}

In Fig.~\ref{sigma plot} we therefore only compare our upper estimate,
obtained by ignoring transverse momenta in the quark propagators and given
by the dot-dashed line, with the data of the combined cross section 
$\sigma(\gamma\gamma\to\pi^+\pi^-,K^+ K^-)$ of Ref.~\cite{Dominick},
where we have accounted for the contributions from kaons by a relative factor
$(f_K/f_\pi)^4\simeq 2.2$. For comparison we again show the collinear 
approximation. As we can see, the curves are already far below the data,
so that the inclusion of $k_\perp$-corrections in the quark propagators
would further increase the discrepancy.

In Ref.~\cite{BrodskyLepage1981} it was shown that, in the collinear 
approximation, essential parts of the hard scattering amplitudes are 
accidentally proportional to the pion form factor. Using this relation and a 
phenomenological value for $F_\pi$ the authors 
of~\cite{BJPR} obtained reasonable agreement with the data. However, we would 
like to emphasise that the assumed value for the pion form factor, 
$Q^2 F_\pi(Q^2)=0.3$ GeV$^2$, is rather large for a perturbative calculation.
With a renormalisation scale of the order of the typical virtuality of the
exchanged gluon and using the asymptotic form of the pion distribution 
amplitude, the pion form factor in the collinear approach 
reads~\cite{LepageBrodsky1980} $Q^2 F_\pi(Q^2)=8 \pi f_\pi \alpha_s(Q^2)$ 
and ranges between 0.17 and 0.1 GeV$^2$ for 1~GeV$\lsim Q \lsim$4~GeV.   
Since the pion form factor enters the cross section for 
$\gamma\gamma\to\pi^+\pi^-$ quadratically that accounts for the 
difference between our result for the collinear approximation and that of
Ref.~\cite{BJPR}.

Finally, we would like to point out that with the inclusion of 
$k_\perp$-corrections in the hard scattering amplitude the simple relation 
between the cross section and the pion form factor does not longer hold. 
In particular, our predictions are independent of any phenomenological value 
for the pion form factor.

\section*{The perturbative limit of the $2\pi$-DA}

We now turn to the kinematical regime where one of the photons has a large 
virtuality $Q^2$. In Refs.~\cite{Mueller} it was shown that for $s\ll Q^2$ the
helicity amplitude of $\gamma^*\gamma\to\pi\pi$ factorises in a hard part and 
a generalised distribution amplitude $\phitwopi$:
\begin{eqnarray}
 {\cal M}_{\lambda\lambda'}(\zeta,s)= \frac{1}{2}\, \delta_{\lambda\lambda'}
  \sum_q\, e_0^2\, e_q^2\,\int_0^1 dz \, \frac{2z-1}{z (1-z)} \,
  \phitwopi^q(z,\zeta,s), \label{helamp} 
\end{eqnarray}
where the light-cone fractions $z=k^+/P^+$ and $\zeta=p^+/P^+$ 
respectively describe how the partons and the pions share the light-cone plus 
component of the total momentum $P=p+p'$ of the pions and the sum runs over all
quark flavours $q$. The $2\pi$-DA, first discussed in~\cite{Grozin}, represents
the collinear hadronisation of two partons into a pion pair. The helicity 
selection rule, expressed through the Kronecker delta, immediately 
follows from the collinear scattering of massless quarks. Note that apart 
from logarithmic corrections the leading order expression~(\ref{helamp}) is 
completely independent of $Q$.

If we demand that $s,-t,-u\gg\Lambda^2$, where $\Lambda$ is a typical hadronic
scale of the order of 1 GeV, while keeping the constraint that the photon
virtuality is the dominant scale, $s\ll Q^2$, we can use the conventional
hard scattering approach~\cite{LepageBrodsky1980} to calculate the helicity 
amplitude~(\ref{helamp}) in terms of the hard scattering amplitude $T_H$ and 
two single pion DAs $\phi_\pi$:
\begin{eqnarray}
 {\cal M}_{\lambda\lambda'}(s,t,u)=\frac{f_\pi^2}{24}\, \int_0^1
  dx\, dy\,\phi_\pi(\bar{y})\,\phi_\pi(x)\,
  T_{H,\lambda\lambda'}(x,y,s,t,u). \label{shsp}
\end{eqnarray}
Using light-cone gauge and organising the result in powers of $\sqrt{s}/Q$
one can show~\cite{DFKV} that the leading contributions are independent of $Q$,
reflecting the correct scaling behaviour, and come from the diagrams
of the group~B in Fig.~\ref{hsp diagrams}. Moreover, the helicity selection 
rule of Eq.~(\ref{helamp}) is reproduced. Reexpressing Eq.~(\ref{shsp})
through the light-cone fractions $z$ and $\zeta$ for each diagram, we can then 
read off the large-$s$ limit of the $2\pi$-DA for a flavour $q=u$ by 
comparison of Eqs.~(\ref{helamp}) and~(\ref{shsp}):
\begin{eqnarray}
 \phitwopi^u(z,\zeta,s)= \frac{8\pi f_\pi^2}{9} 
  \Bigg\{&&\hspace{0mm}\Theta(\zeta-z)\, \frac{\zeta}{\zeta-z} \, 
  \phi_\pi\bigg(\frac{z}{\zeta}\bigg) \; I(\bar{z},\bar{\zeta},s;\phi_\pi) 
\nonumber\\
  -&&\hspace{0mm}\Theta(z-\zeta)\, \frac{\bar{\zeta}}{z-\zeta} \,
   \phi_\pi\bigg(\frac{\bar{z}}{\bar{\zeta}}\bigg) \; 
   I(z,\zeta,s;\phi_\pi) \Bigg\}, \label{model DA}
\end{eqnarray}
where the integral $I$ is given by
$   I(z,\zeta,s;\phi_\pi)=\int_0^1 dx \, \frac{\alpha_s}{s} \, 
    \frac{z+\bar{x}\zeta}{z-x\zeta} \, \frac{\phi_\pi(x)}{\bar{x}}\,.
$
The $2\pi$-DAs for u- and d-quarks are related by 
$\phitwopi^u(z,\zeta,s)=-\phitwopi^{d}(\bar{z},\zeta,s)$ and since higher
Fock states are suppressed by powers of $\alpha_s/s$ there is no s-quark
contribution. The $1/s$ scaling of Eq.~(\ref{model DA}) is a characteristic
feature of the hard scattering 
approach~\cite{LepageBrodsky1980,BrodskyLepage1981}.

Our result manifestly fulfills the charge conjugation relation
$\phitwopi^q(z,\zeta,s)=-\phitwopi^q(\bar{z},\bar{\zeta},s)$ and it can  
be shown to comply with a general polynomiality condition~\cite{Polyakov1999}.
It possesses integrable logarithmic singularities at $z=\zeta$, which 
reflect the above mentioned endpoint problems of the collinear hard scattering
approach when the exchanged gluon becomes soft.

\section*{Conclusions}

Two-photon annihilation into pion pairs allows for a sensitive test of 
perturbative QCD. Using a self-consistent approach, where there are no large 
endpoint contributions spoiling the applicability of perturbation theory, we 
have shown that the hard contributions are not sufficient to explain the 
experimental data of $\gamma\gamma\to\pi^+\pi^-$. Therefore considerable soft 
contributions have to be expected. New data are desireable to determine the 
onset of the perturbative regime, which seems not to start below c.m. energies
of 4-5 GeV.
When one of the photons is far off-shell and at large $s$, where transverse 
momenta become irrelevant, the collinear hard scattering approach can be 
applied to calculate the perturbative limit of the $2\pi$-DA in terms of the 
conventional pion distribution amplitudes.

\section*{acknowledgements}

I would like to thank M. Diehl, Th. Feldmann and P. Kroll for a fruitfull 
collaboration, on which the second part of the talk is based,
and R. Jakob for critical remarks. It is a pleasure
to thank A. Finch and co-workers for an excellent organisation
of the conference. A graduate grant by the Deutsche Forschungsgemeinschaft is
acknowledged.

\end{document}